# Unfolding the effects of final-state interactions and quantum statistics in two-particle angular correlations

Łukasz Kamil Graczykowski * and Małgorzata Anna Janik †
*Faculty of Physics, Warsaw University of Technology ul. Koszykowa 75, 00-662 Warszawa, Poland*

Angular correlations of identified particles measured in ultrarelativistic proton-proton (pp) and heavy-ion collisions exhibit a number of features which depend on the collision system and particle type under consideration. Those features are produced by mechanisms, such as (mini)jets, elliptic flow, resonance decays, and conservation laws. In addition, of particular importance are those related to the quantum statistics (QS) and final-state interactions (FSIs). In this paper we show how to unfold the QS and FSI contributions in angular correlation functions by employing a Monte Carlo approach and using momentum correlations (femtoscopy), focusing on pp reactions. We validate the proposed method with PYTHIA 8 Monte Carlo simulations of pp collisions at $\sqrt{s} = 13$ TeV coupled to calculations of QS and FSI effects with the Lednický and Lyuboshitz formalism and provide predictions for the unfolded effects. In particular, we show how those effects modify the shape of the angular correlation function with emphasis on pions and protons. Most importantly, specific structures observed in the near-side region for both baryon-baryon and baryon-antibaryon pairs, originating from the strong interaction, are unveiled with the proposed method.

## I. INTRODUCTION

Collisions of ultrarelativistic protons and heavy ions allows one to study the quantum chromodynamics (QCD) with unprecedented precision. Various tools can be used to gain insight into the underlying processes governing them. In this paper we focus on angular correlations which can be employed to access the particle production mechanism.

Two-particle correlations measured as a function of relative pseudarapidity $\Delta\eta = \eta_1 - \eta_2$ and azimuthal angle $\Delta\varphi = \varphi_1 - \varphi_2$ (where indices "1" and "2" denote the two particles of the pair) are referred to as the angular correlations. They are sensitive to a number of physical mechanisms, such as (mini)jets, elliptic flow, Bose-Einstein or Fermi-Dirac quantum statistics (in the case of identical bosons or fermions), resonance decays, conservation laws, etc., allowing to study both short- (limited in phase space) and long-range (event wide) effects. Each of these mechanisms manifests differently in the global angular correlation picture. However, a typical correlation function, observed in proton-proton (pp) collisions, can be characterized by the presence of a near-side peak, an away-side ridge, and an underlying correlation originating from the momentum conservation. The influence of those mechanisms which manifest in elementary particle collisions is schematically shown in Fig. 1.

Studies of pp collisions at $\sqrt{s} = 7$ TeV performed for identified hadrons by the ALICE Collaboration at the Large Hadron Collaboration revealed that the global correlation shape changes for different particle species [1]. In particular, correlations of baryon-baryon pairs (combined with antibaryon-antibaryon ones) in pp collisions qualitatively differ from the correlations of two mesons and baryon-antibaryon pairs. They exhibit only a depletion around $(\Delta\varphi, \Delta\eta) \approx (0, 0)$, similar to correlation arising from the momentum conservation only and lack typical near- and away-side structures originating mainly from (mini)jets. The main conclusion from Ref. [1] is that the correlation pattern remains very similar for $\mathrm{pp} \oplus \overline{\mathrm{pp}}$, $\mathrm{p}\Lambda \oplus \overline{\mathrm{p}}\overline{\Lambda}$, and $\Lambda\Lambda \oplus \overline{\Lambda}\overline{\Lambda}$ pairs. Thus, the final shape of the correlation function may be connected to the intrinsic nature of baryon production. Moreover, none of the observed two-baryon correlations agree even qualitatively with PYTHIA [2] (various tunes of versions 6.4 [3,4] and 8 [5,6]) and PHOJET [7] simulations with models predicting a correlation with jetlike near- and away-side structures for those pairs.

The ALICE Collaboration results have been later followed up by STAR Collaboration measurements in Au-Au collisions in the beam energy scan (BES) program at the Brookhaven National Laboratory Relativistic Heavy Ion Collider [8]. In that case, the two-particle correlation was measured as a function of rapidity difference $\Delta y = y_1 - y_2$, instead of $\Delta\eta$. In particular, the near-side depletion is also visible in Au-Au collisions across BES collision energies [8]. However, for $\sqrt{s_{\mathrm{NN}}} = 200$-GeV data, it is convoluted with the underlying

*Corresponding author: lukasz.graczykowski@pw.edu.pl
†malgorzata.janik@pw.edu.pl

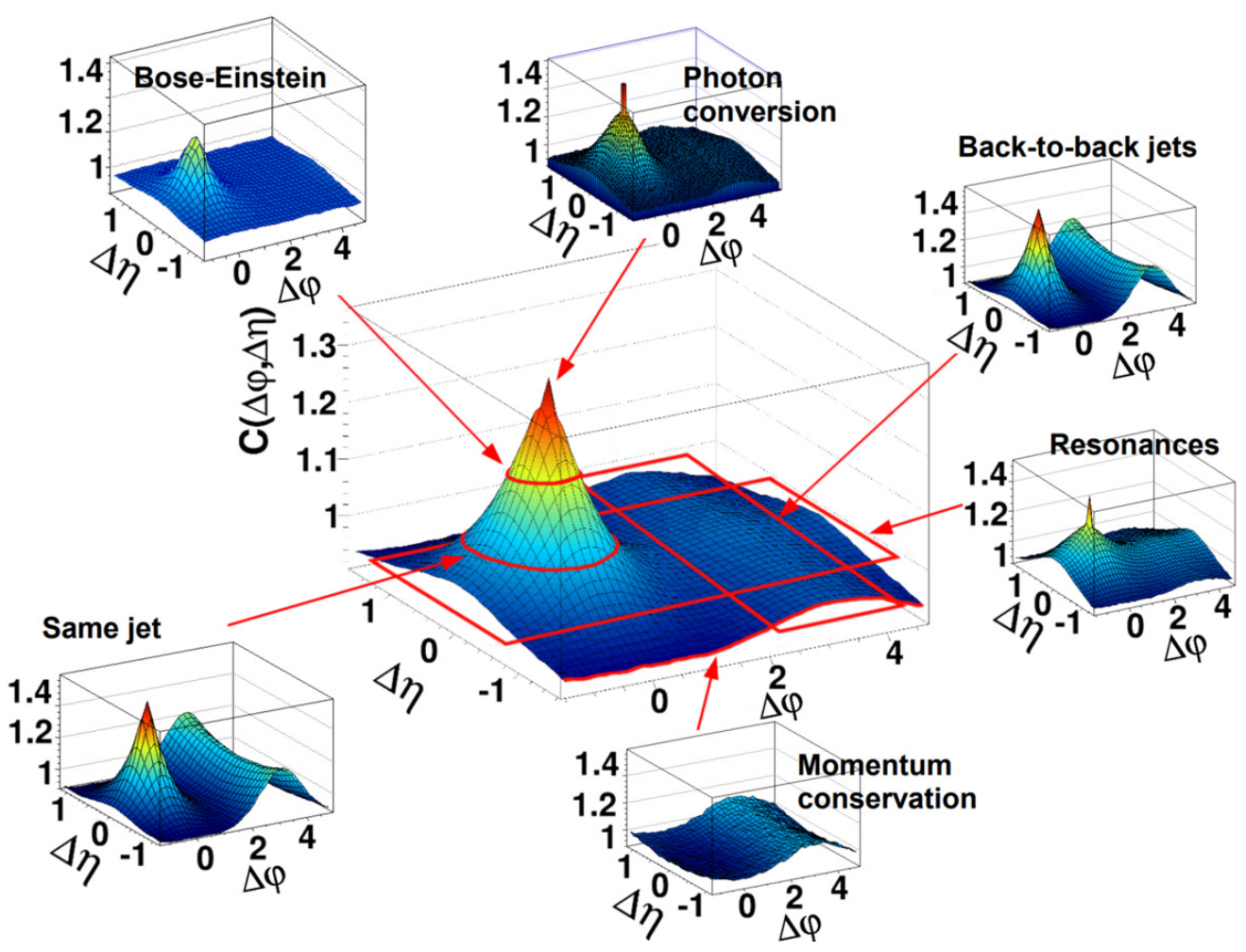

FIG. 1. Schematic of the angular correlation function $C(\Delta\eta, \Delta\varphi)$ showing contributions from various correlation sources.

flow correlation, which produces specific near- and away-side ridges.

Recent theoretical developments suggest that the quark coalescence process may be responsible for the presence of this effect in ALICE Collaboration and STAR Collaboration data [9,10], whereas others suggest that we are lacking some fundamental insight into baryon production in the string model [11].

ALICE Collaboration and STAR Collaboration results reveal yet another interesting observation. In two-proton correlations an additional small peak convoluted with the depletion at exactly $(\Delta\varphi, \Delta\eta) = (0, 0)$ is visible in ALICE Collaboration data, whereas the $p\bar{p}$ correlation from the STAR Collaboration also exhibits a depletion around $(\Delta\varphi, \Delta\eta) \approx (0, 0)$, although much narrower with respect to the baryon-baryon case. Both structures, the small peak in pp and the depletion in $p\bar{p}$ correlations, are postulated to originate from the strong final-state interaction (FSI). Verification of this hypothesis is one of the goals of this paper.

The strong final-state interaction in relativistic particle collisions is studied by using the technique of femtoscopy, that is, the measurement of two-particle correlations as a function of the pair relative momentum [12]. In recent years this technique has proven to be a robust tool allowing for a precise determination of the strong FSI for various hadron-hadron pairs both in elementary and heavy-ion collisions, see, i.e., Refs. [13,14].

Both angular and femtoscopic correlation functions are measured experimentally. The effects of quantum statistics (QS) and FSI are well studied and understood in the case of the femtoscopic representation; however, their contribution to the angular correlation function is less clear. In this paper we propose a Monte Carlo procedure that links the angular correlation function with momentum correlations and employs the femtoscopic formalism to unfold the strong interaction component in the angular space. We demonstrate and validate the procedure with PYTHIA simulations, coupled to the Lednický and Lyuboshitz [15] formalism for the calculation of the QS and FSI effects [15–17] and present how the strong interaction component manifests in the angular space for both baryon-baryon and baryon-antibaryon cases. For reference, we also study correlations of same and opposite-charge pion pairs where the Bose-Einstein QS and Coulomb FSI are significant ingredients, respectively.

## II. DEFINITION OF TWO-PARTICLE CORRELATION FUNCTION

In general, the two-particle correlation function (CF) can be defined as a ratio of the conditional probability of observing two particles simultaneously, divided by the product of probabilities of observing them independently,

$$C(\mathbf{p_1}, \mathbf{p_2}) = \frac{P_{1,2}(\mathbf{p_1}, \mathbf{p_2})}{P_1(\mathbf{p_1}) \cdot P_2(\mathbf{p_2})}, \quad (1)$$

where $\mathbf{p_1}$ and $\mathbf{p_2}$ denote momenta of the first and second particles in the pair, respectively [18]. Such a definition is valid for any representation of the two-particle CF [18].

In the experiment the CF is then defined as a ratio of distribution of pairs constructed from the same collision ("same events"), usually referred to as the "signal distribution, divided by a distribution of pairs of particles where each particle comes from a different event, usually referred to as the "background distribution," which is the basis of the so-called "event mixing" technique [12].

In order to conveniently analyze the CF defined in Eq. (1), calculations are performed in various representations, depending on the physics mechanism under study, which leads to the reduction of dimensions in data processing. For instance, the angular correlations are expressed as a function of $\Delta\eta$ and $\Delta\varphi$, and femtoscopic correlations are frequently expressed as a function of the magnitude of the half of the pair relative momentum $k^* = |\mathbf{k}^*|$, which is equal to the first particle's momentum in the pair rest frame (PRF) where the pair center of mass is at rest ($\mathbf{p_1} = -\mathbf{p_2}$).

In general, each physics mechanism being a correlation source influences all representations; however, different effects are pronounced to the varying degree depending on the representation. For example, those related to QS and FSI and studied using the technique of femtoscopy in the momentum representation also exhibit specific shapes in the angular representation. Similarly, jets are frequently studied in the angular correlation representation, but they are also present as a prominent background with a dependence on $k^*$ in the femtoscopic representation.

In this paper we concentrate on the effects of QS and FSI, and via the femtoscopic formalism investigate their contribution to the overall angular correlation shape.

## III. DESCRIPTION OF DATA SIMULATION

### A. Choice of the Monte Carlo model

In order to perform calculations proposed in this paper, an event generator is needed. In elementary particle collisions, such as pp, PYTHIA [2] is one of the most successful general-purpose models used to calculate the dynamics of particle collision at ultrarelativistic energies. Therefore, it is a natural choice for this paper. It includes perturbative QCD calculations for interactions at high momentum transfer and phenomenological models for the description of the processes in the low-momentum regime. For modeling of the hadronization process the Lund string fragmentation approach [19] is employed. The Bose-Einstein QS effects are implemented in PYTHIA [20,21] as well, however, by default, they are switched off [5]. In this paper a more complete description of QS and FSI is employed as discussed in Sec. III B, and, therefore, this option is not used.

In this paper PYTHIA version 8.4 [5] Monash tune [6] is used for simulation of pp collisions at $\sqrt{s} = 13$ TeV center-of-mass energy. A total number of $12 \times 10^8$ minimum-bias events were simulated and used for calculation of the correlation functions.

### B. QS and FSI afterburners

None of the Monte Carlo event generators includes a complete description of the effects related to QS and FSI [22,23]. Hence, so-called "femtoscopic afterburners" are used to introduce them *a posteriori*.

As described in Ref. [18], in theoretical modeling probabilities in Eq. (2) can be expressed using respective single- and two-particle emission functions. For instance, probability to emit a particle pair from given space-time points with a given momentum can be expressed using the emission function $S(\mathbf{k}^*, \mathbf{r}^*)$, where $\mathbf{r}^* = \mathbf{x_1} - \mathbf{x_2}$ is the separation vector between space-time emission points $\mathbf{x_1}$ and $\mathbf{x_2}$ of the first and second particles in the pair, respectively. It is also referred to as the "source function." In general, it should contain all physics processes, including those related to QS and FSI; however, since models do not account for them one can assume their independence from the emission process. This yields the following form of the CF in the momentum representation [24,25]:

$$C(\mathbf{k}^*) = \int S(\mathbf{k}^*, \mathbf{r}^*)\, |\Psi(\mathbf{k}^*, \mathbf{r}^*)|^2 d^4\mathbf{r}^*, \tag{2}$$

where $S(\mathbf{k}^*, \mathbf{r}^*)$ is the two-particle source emission function and $\Psi(\mathbf{k}^*, \mathbf{r}^*)$ is the pair wave function. The pair wave function depends on the QS and FSI effects between particles forming pairs under consideration and their kinematic properties [15,16]. Therefore, for each pair, the modulus square of the pair wave function is calculated as an additional weight. In the case of the equal emission times of both particles in PRF it is represented by the Bethe-Salpeter amplitude which coincides with a stationary solution of the scattering problem with reversed time direction in the emission process [17]. The mathematical procedure of calculating such weights, referred to as the Lednický and Lyuboshitz [15] formalism, is described in Appendix.

PYTHIA does not provide the space-time emission points of final-state particles; therefore, the form of the $S(\mathbf{k}^*, \mathbf{r}^*)$ has to be introduced by other means. In this paper we assume a simplified spherically symmetric Gaussian distribution in the PRF, $S(\mathbf{r}^*) = \frac{1}{(4\pi R_{\mathrm{inv}}^2)^{3/2}} \exp(-\mathbf{r}^{*2}/4R_{\mathrm{inv}}^2)$, where $R_{\mathrm{inv}} = 1.5$ fm is the source size. This is based on the ALICE Collaboration observations from femtoscopic studies in pp collisions at $\sqrt{s} = 7$ and 13 TeV [26]. We note that the non-Gaussian forms of the source have been proposed and successfully employed in the analysis of experimental data analysis [27–32]. However, the proposed procedure is independent of this ansatz.

## IV. CORRELATION FUNCTION CALCULATION

The procedure of obtaining the CF from the simulated data resembles the approach performed in the experiment which has been introduced in Sec. II; however, in order to distinguish the components of QS and FSI from the underlying mechanisms in the simulations, we define the following three variants of the signal distribution. (1) The first distribution $S$ is created when each same-event pair is added with the same weight equal to 1.0. (2) The second distribution $S_{\mathrm{w}}$ is created when each same-event pair is added with a weight equal to the Bethe-Salpeter amplitude, which is calculated according to the procedure described in Sec. III B. (3) The third distribution $M_w$ is calculated from mixed-event pairs, where each pair is added to the distribution also with a weight calculated as above. Subsequently, three distinct types of the CF can be constructed, each containing different information:

(1) $C_{\mathrm{base}} = S/M$, where $M$ is the mixed-event distribution, contains only the event-wide correlations without the QS and FSI effects added by the afterburner;
(2) $C_{\mathrm{full}} = S_{\mathrm{w}}/M$ contains the full information, that is, the event-wide correlations with additional effects of QS and FSI added by the afterburner;
(3) $C_{\mathrm{QS+FSI}} = M_{\mathrm{w}}/M$ contains only the effects related to QS and FSI and is an equivalent to numerical integration of Eq. (2).

In this paper, all three CFs are expressed as a function of either $k^*$ or $\Delta\eta$ and $\Delta\varphi$ in the femtoscopic or angular representation, respectively. Examples of all three correlation functions in both representations, for $\pi^+\pi^+$, $\pi^+\pi^-$, pp, and $\mathrm{p\overline{p}}$ pairs, calculated from PYTHIA 8 simulated data for pp collisions at $\sqrt{s} = 13$ TeV are shown in Fig. 2. No differences other than statistical fluctuations are expected for the corresponding charge conjugate pairs (i.e., $\pi^-\pi^-$, $\mathrm{\overline{p}\overline{p}}$).

## V. UNFOLDING PROCEDURE

In Monte Carlo simulations we can easily switch on or off the QS and FSI afterburners. Consequently, for those particle pairs where these effects are well known in theory and the afterburner produces meaningful results, we can investigate their contribution to the overall shape of the angular correlation by defining CFs as in the previous section. However, this approach will not work for those pairs where the strong component of the FSI is poorly known or not known at all. In this section we propose and describe the algorithm of the procedure allowing to unfold the effects of QS and FSI, which are typically studied with femtoscopy, in angular correlations. This allows to use experimentally measured data without the need of relying on simulations.

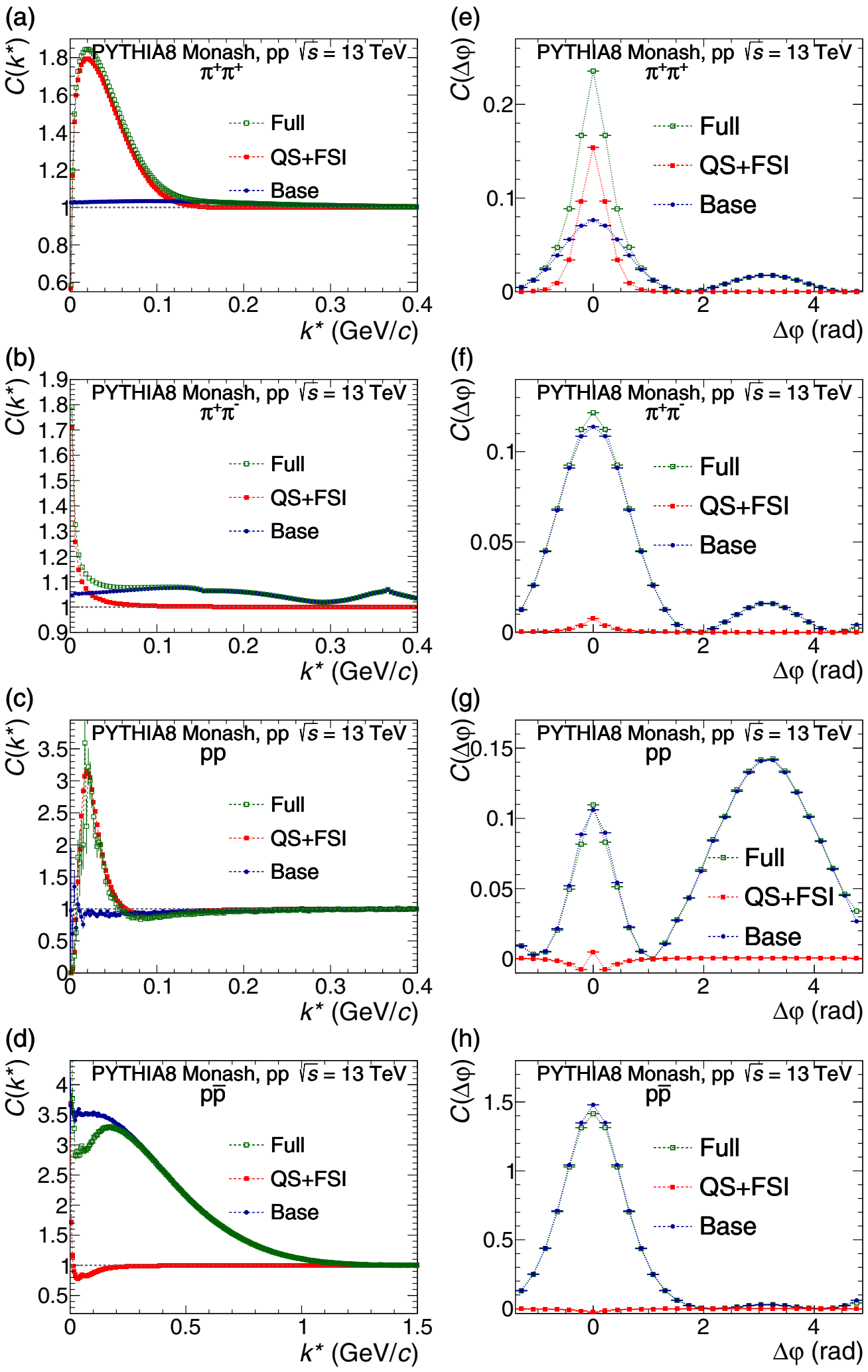

FIG. 2. Correlation functions in (left) femtoscopic and (right) angular representations of $\pi^+\pi^+$ pairs (first row), $\pi^+\pi^-$ pairs (second row), pp pairs (third row), and $p\bar{p}$ pairs (fourth row) from PYTHIA 8 simulations, coupled to the Lednický and Lyuboshitz [15] formalism of pp collisions at $\sqrt{s} = 13$ TeV, showing $C_{\text{full}}$ (in green), $C_{\text{QS+FSI}}$ (in red), and $C_{\text{base}}$ (in blue). For better visualization of the differences, the $\Delta\varphi$ projections are shifted by a constant value so that the baseline is at $C(\Delta\varphi) = 0$.

The proposed procedure employs a Monte Carlo approach of creating particles from sampling single-particle kinematic distributions, such as the transverse momentum ($p_T$), pseudorapidity ($\eta$), and azimuthal angle ($\varphi$), which are normally measured in the experiment. Each particle is then represented by a set of three numbers. Then, they are combined in pairs, and each pair is assigned a weight from the experimentally measured femtoscopic correlation function. Such an approach allows for studies of the short-range correlations, such as the effects of QS and FSI; however, it is not sensitive to long-range phenomena, i.e., the energy-momentum conservation, which is by definition not present while constructing pairs in that simple way.

*Input.* The main required ingredient for the procedure is a measured femtoscopic correlation function $C(k^*)$ for a given particle pair under consideration. In addition, single-particle experimental distributions of $p_T$, $\eta$, and $\varphi$ are needed. Alternatively, since collider experiments are designed to have a uniform acceptance in $\eta$ and $\varphi$, it should be possible to substitute the $\eta$ and $\varphi$ distributions by uniform ones without introducing significant distortions to the results.

*Algorithm flow.* In order to create particles from the input data one needs to start from sampling the $p_T$, $\eta$, and $\varphi$ distributions provided in the input. Each particle will be described then by a set of three pseudorandom values of $p_T$, $\eta$, and $\varphi$. Next, particles can be combined in pairs and for each pair one can easily calculate $\Delta\eta$ and $\Delta\varphi$ corresponding to the given pair. In order to obtain the background distribution $B(\Delta\eta, \Delta\varphi)$, each pair is added with the same weight. In order to obtain the signal distribution, for each pair a weight $w = C(k^*)$ from the experimental femtoscopic correlation function has to be extracted. Therefore, for each pair the value of the pair relative momentum $k^*$ has to be calculated from sampled kinematic quantities of a given pair. Then, the pair is added to the the signal distribution $A(\Delta\eta, \Delta\varphi)$ with a given weight $w$. Finally, the angular correlation unfolded from the femtoscopic correlation function is $C^{\text{unfolded}}(\Delta\eta, \Delta\varphi) = A(\Delta\eta, \Delta\varphi)/B(\Delta\eta, \Delta\varphi)$. The $C^{\text{unfolded}}(\Delta\eta, \Delta\varphi)$ distribution obtained in the way described above can be then used for direct comparisons of angular correlation functions obtained in the experiment.

One should note that the method is dependent on the $k^*$ range which is taken into account in the $C(k^*)$ femtoscopic correlation function. The femtoscopic region (small values of $k^*$), corresponds to a limited phase space. As already explained, the proposed method will not be sensitive to the event-wide correlation effects that manifest at large $k^*$ values. Figure 3 presents how $k^*$ ranges in the femtoscopic correlation functions correspond to given $\Delta\eta\Delta\varphi$ regions. One can clearly see that the femtoscopic region clearly translates to the near-side region of the angular CF, which is the subject of our interest.

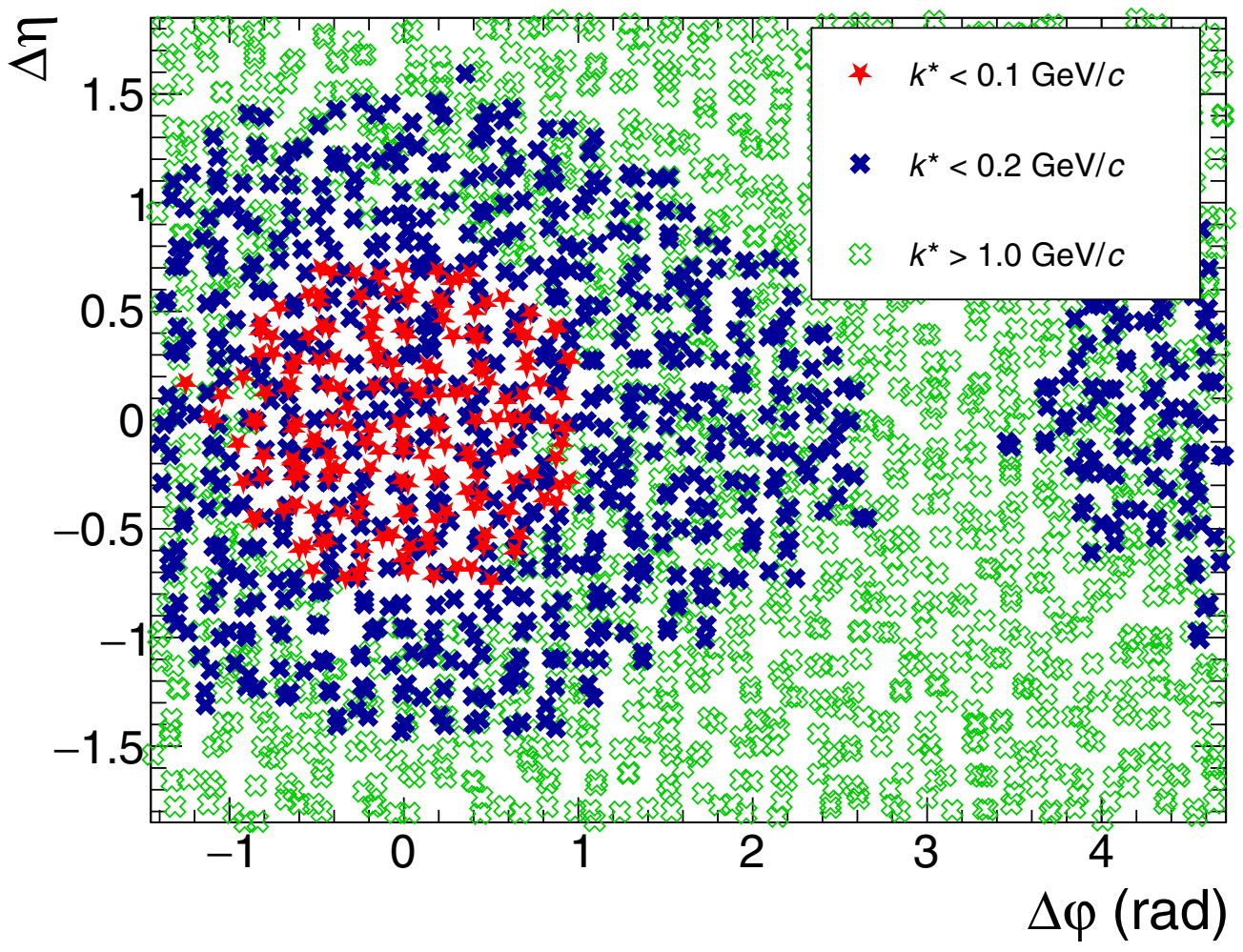

FIG. 3. Schematic showing which $k^*$ ranges in the femtoscopic correlation function corresponds to which $\Delta\eta\Delta\varphi$ regions in the angular correlation.

## VI. RESULTS AND DISCUSSION

This paper focuses on same and opposite charge pair combinations of pions and protons. The correlation functions for all four pair combinations in the femtoscopic representation $C_{\mathrm{QS+FSI}}(k^*)$ are shown in the left panels of Fig. 2. The effects of QS and FSI variously manifest in those pairs. In the case of $\pi^+\pi^+$ pairs the most dominant effect is the Bose-Einstein QS, reflected by the peak structure at low $k^*$, which is sensitive to the size of the pion emitting source [12]. Additionally, effects related to the Coulomb and strong FSI are present with the former affecting the correlation function for the few first $k^*$ bins for both pion and protons pairs, and the latter having an almost negligible effect for pion pairs. In the case of $\pi^+\pi^-$ pairs, the correlation effect is driven essentially by the Coulomb FSI only [12]. Similarly the strong and Coulomb FSI also play a significant role [17] for pp pairs. In particular, the strong interaction component produces a characteristic peak with a maximum around $k^* \approx 50\ \mathrm{MeV}/c$. Such sensitivity allows for a precise study of the strong interaction between various baryon-baryon pairs; see, i.e., Refs. [13,14]. In terms of QS the pp pairs also the Fermi-Dirac quantum statistics is present. The $p\bar{p}$ correlations are sensitive to the strong interaction as well, especially to its inelastic channel, allowing for the determination of the annihilation process [33]. Therefore, these four pair combinations can be used to quantify individually the QS and FSI effects in the angular representation and allow for a validation of the unfolding procedure.

The QS + FSI correlation functions in the angular representation for all four pair combinations are presented in Fig. 4. Panels in the left column show the $C_{\mathrm{QS+FSI}}(\Delta\eta, \Delta\varphi)$ correlations obtained directly from the PYTHIA 8 simulation with the QS + FSI afterburner. Panels in the middle column show the unfolded correlations $C^{\mathrm{unfolded}}_{\mathrm{QS+FSI}}(\Delta\eta, \Delta\varphi)$, obtained from $C_{\mathrm{QS+FSI}}(k^*)$ femtoscopic correlations from the PYTHIA 8 simulations with the QS + FSI afterburner, according to the procedure described in Sec. V. The ratios $C^{\mathrm{unfolded}}_{\mathrm{QS+FSI}}(\Delta\eta, \Delta\varphi)/C_{\mathrm{QS+FSI}}(\Delta\eta, \Delta\varphi)$ for each pair are shown in panels in the right column. The $\Delta\eta$ and $\Delta\varphi$ projections of these distributions are shown in Fig. 5. From the correlation functions of $\pi^+\pi^+$ pairs one can clearly see that the Bose-Einstein QS produces a significant peak around $(\Delta\eta, \Delta\varphi) \approx (0, 0)$ with a magnitude around 1.4. On the other hand, the Coulomb FSI effect, observed for the $\pi^+\pi^-$ correlations is much less pronounced, reaching the magnitude of 1.03. In the case of pp pairs a narrow depletion around $(\Delta\eta, \Delta\varphi) \approx (0, 0)$ can be observed with an additional narrow peak located exactly at $(\Delta\eta, \Delta\varphi) = (0, 0)$. This result is in line with the experimental analysis by the ALICE Collaboration [34], which also observed a depletion convoluted with an additional small peak for pp pairs. Although the depletion in the experimental analysis is a result of the wide-range energy-momentum conservation, the origin of the small peak was postulated to arise from the two-proton strong interaction. This paper proves that assumption.

Proton-antiproton pairs are also interesting as another depletion is visible in that case. It originates from the anticorrelation structure seen in the femtoscopic correlation function [red line of panel (d) in Fig. 2] which is produced by the inelastic channel of the $p\bar{p}$ strong interaction (annihilation). That structure is also preserved in the unfolded correlation. Therefore, this paper validates with a detailed model simulations, the strong interaction hypothesis for both experimentally observed structures in the ALICE Collaboration and the STAR Collaboration.

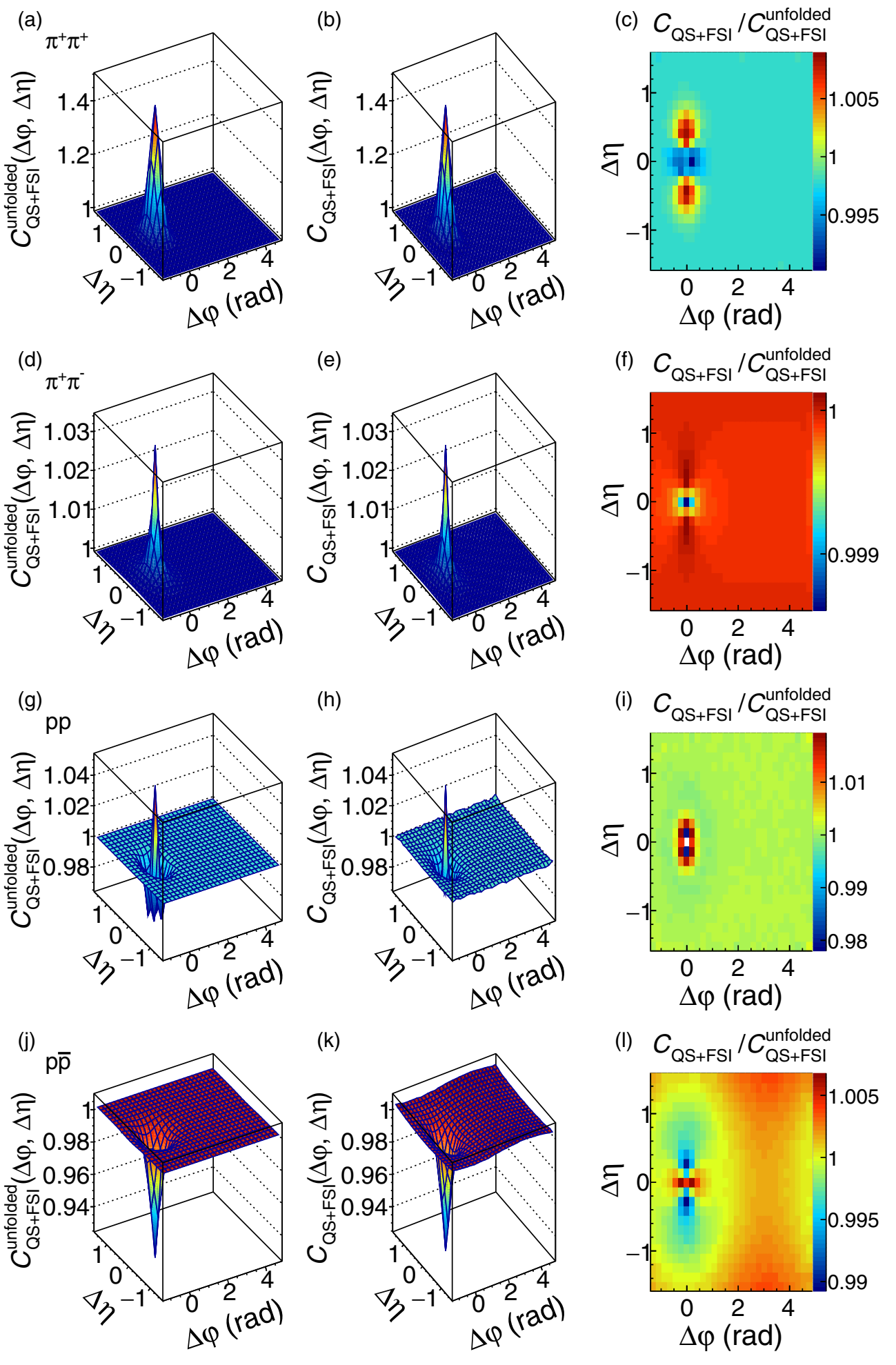

FIG. 4. Correlation functions $C^{\mathrm{unfolded}}_{\mathrm{QS+FSI}}(\Delta\eta, \Delta\varphi)$ (left column) and $C_{\mathrm{QS+FSI}}(\Delta\eta, \Delta\varphi)$ (middle column) in the angular representation of $\pi^+\pi^+$, $\pi^+\pi^-$, pp, and $p\bar{p}$ pairs from PYTHIA simulated pp collisions, coupled to the Lednický and Lyuboshitz [15] formalism at $\sqrt{s} = 13$ TeV. The right column shows the ratio of the ratio $C_{\mathrm{QS+FSI}}(\Delta\eta, \Delta\varphi)/C^{\mathrm{unfolded}}_{\mathrm{QS+FSI}}(\Delta\eta, \Delta\varphi)$.

Accuracy of the procedure can be quantified by comparing the $C^{\mathrm{unfolded}}_{\mathrm{QS+FSI}}(\Delta\eta, \Delta\varphi)$ correlations with the ones obtained directly from PYTHIA 8 simulations with the QS and FSI afterburners $C_{\mathrm{QS+FSI}}(\Delta\eta, \Delta\varphi)$. The shape of the near-side peak is well reproduced by the unfolding procedure for same and opposite charge pion pairs with relative differences smaller than 8% and 5%, respectively. In the case of pp pairs, the unfolding procedure does reproduce qualitatively both observed features, that is, the depletion centered around $(\Delta\eta, \Delta\varphi) \approx (0, 0)$ with a narrow peak located exactly at $(\Delta\eta, \Delta\varphi) = (0, 0)$, although with a different magnitude and slightly different shape. In the $p\bar{p}$ system, the depletion around $(\Delta\eta, \Delta\varphi) \approx$ (0, 0) is also reproduced by the unfolding procedure. However, since the $\mathrm{p\bar{p}}$ strong interaction has a wider dependence in $k^*$, its contribution to the $(\Delta\eta, \Delta\varphi)$ angular correlation function is not limited to the $\approx(0, 0)$ region only. Nevertheless, the distinct depletion observed in the near-side region in that case is reproduced well with relative differences reaching a maximum of 30%.

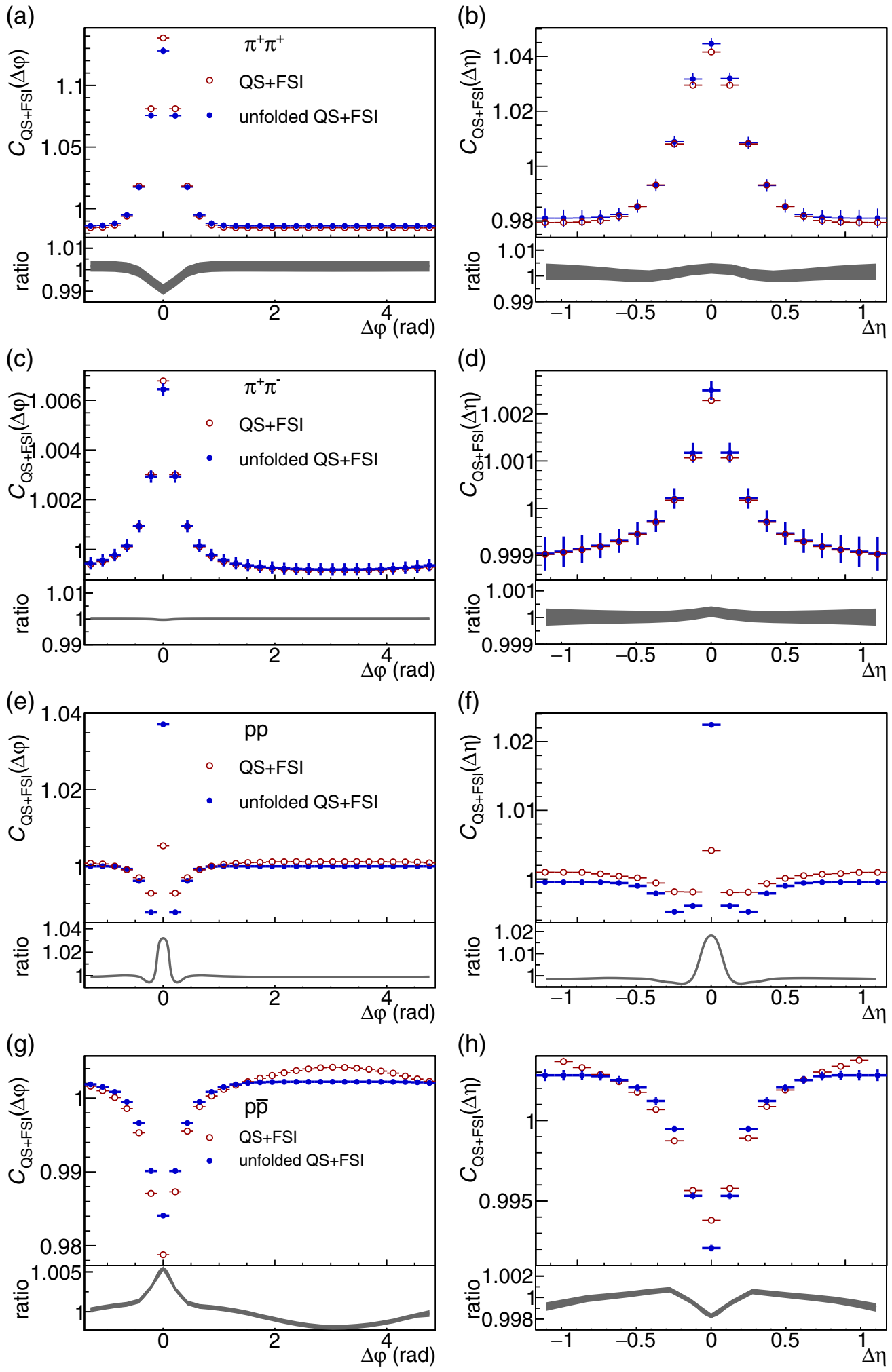

FIG. 5. Upper panels: projections of $C_{\mathrm{QS+FSI}}(\Delta\eta, \Delta\varphi)$ (closed symbols) and $C^{\mathrm{unfolded}}_{\mathrm{QS+FSI}}(\Delta\eta, \Delta\varphi)$ (open symbols) in $\Delta\varphi$ for $|\Delta\eta| < 1.3$ (left column) and in $\Delta\eta$ for $-\frac{\pi}{2} < \Delta\varphi < \frac{\pi}{2}$ (right column) for $\pi^+\pi^+$, $\pi^+\pi^-$, pp, and $\mathrm{p\bar{p}}$ pairs from PYTHIA simulated pp collisions, coupled to the Lednický and Lyuboshitz [15] formalism, at $\sqrt{s} = 13$ TeV. Lower panels: ratios of $C^{\mathrm{unfolded}}_{\mathrm{QS+FSI}}$ to $C_{\mathrm{QS+FSI}}$.

The $\Delta\eta$ and $\Delta\varphi$ projections of $C_{\mathrm{base}}(\Delta\eta, \Delta\varphi)$ and $C_{\mathrm{full}}(\Delta\eta, \Delta\varphi)$ correlations for $\pi^+\pi^+$ pairs are shown in Figs. 6 and 7, respectively. We can clearly see that the shape of the unfolded correlation disagrees with the one calculated directly from the model. The reason is that naturally the effects originating from jets and the energy-momentum conservation occupy a wide range of the phase space. Since the unfolding procedure is based on sampling single-particle distributions, such effects are not accounted for. Therefore, part of the information on the wide-range correlations is lost in the unfolding procedure, leading to poor description of the signal in such case. This is also the reason for the less accurate result of the unfolding the $\mathrm{p\bar{p}}$ $C_{\mathrm{QS+FSI}}(\Delta\eta, \Delta\varphi)$ correlations due to the wide-range effect of the annihilation in $k^*$ as seen in panel (d) of Fig. 2. The unfolded correlations are then more diluted, and the away-side structure is not present at all. However, from Fig. 7 where all short- and wide-range effects are present, we can certainly quantify that the description of the near-side structure is preserved, concludin that the Bose-Einstein QS effect is the dominant effect for $\pi^+\pi^+$ pairs and has to be accounted whereas comparing the experimental results to model calculations.

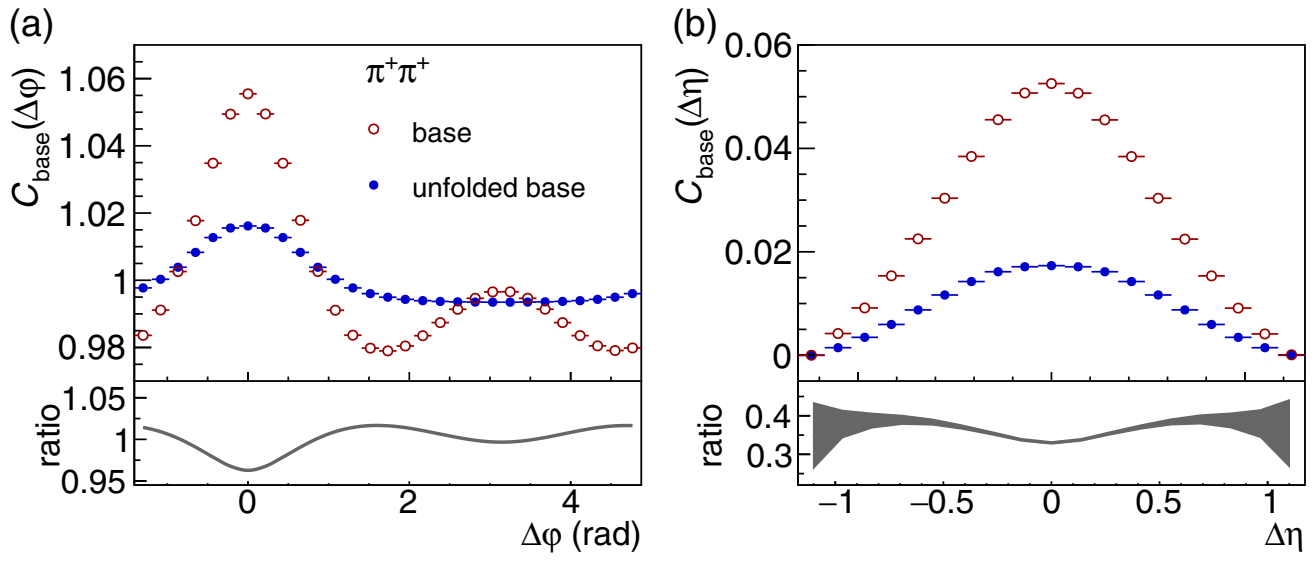

FIG. 6. Upper panels: projections of $C_{\mathrm{base}}(\Delta\eta, \Delta\varphi)$ (closed symbols) and $C^{\mathrm{unfolded}}_{\mathrm{base}}(\Delta\eta, \Delta\varphi)$ (open symbols) in $\Delta\varphi$ for $|\Delta\eta| < 1.3$ (left column) and in $\Delta\eta$ for $-\frac{\pi}{2} < \Delta\varphi < \frac{\pi}{2}$ (right column) for $\pi^+\pi^+$ pairs from PYTHIA simulated pp collisions, coupled to the Lednický and Lyuboshitz [15] formalism at $\sqrt{s} = 13$ TeV. For better visualization of the differences, the $\Delta\eta$ projections are shifted by a constant value so that the baseline is at $C(\Delta\eta) = 0$. Lower panels: ratio of $C^{\mathrm{unfolded}}_{\mathrm{base}}$ to $C_{\mathrm{base}}$.

## VII. SUMMARY

In this paper a Monte Carlo procedure allowing to unfold the effects of quantum statistics and final-state interactions in the angular correlation functions from the femtoscopic measurements is introduced. It was validated with dedicated PYTHIA 8 simulations of pp collisions at $\sqrt{s} = 13$ TeV, coupled to Lednický and Lyuboshitz [15] formalism. Correlation functions in the femtoscopic and angular representations were calculated for the same and opposite charge pairs of pions and protons. The unfolded angular correlations are compared to the ones obtained directly from the simulation, allowing for validation of the proposed procedure. The results show that the unfolding works very well for both $\pi^+\pi^+$ and $\pi^+\pi^-$ pairs in describing the Bose-Einstein QS and Coulomb FSI components of the angular correlation function with relative differences less than 8% and 5%, respectively. Nontrivial structures are observed for pp and $\mathrm{p\bar{p}}$ pairs. In the former case, a depletion in the correlation function around $(\Delta\eta, \Delta\varphi) \approx (0, 0)$ is visible with an additional peak structure directly at $(\Delta\eta, \Delta\varphi) = (0, 0)$. Although the magnitude of the peak is substantially smaller in the unfolded correlation, the procedure is able to describe the shape qualitatively. A qualitatively similar peak structure, located at $(\Delta\eta, \Delta\varphi) = (0, 0)$, in the middle of the depletion, was observed experimentally by the ALICE Collaboration and postulated to result from the strong two-proton interaction. This paper validates this ansatz. In the $\mathrm{p\bar{p}}$ case, another depletion around $(\Delta\eta, \Delta\varphi) \approx (0, 0)$ is revealed, originating from the annihilation process. However, due to the fact that the $\mathrm{p\bar{p}}$ strong interaction has a wide dependence in $k^*$, it contributes not only to the $(\Delta\eta, \Delta\varphi) \approx (0, 0)$ region of the angular correlation function. Nevertheless, the specific depletion observed in the near-side region in that case is reproduced with relative differences reaching a maximum of 30%. The depletion observed in the $\mathrm{p\bar{p}}$ correlations is qualitatively in agreement with the experimental observations by the STAR Collaboration in Au-Au collisions from the Beam Energy Scan Program also confirming the postulation of its $\mathrm{p\bar{p}}$ annihilation origin.

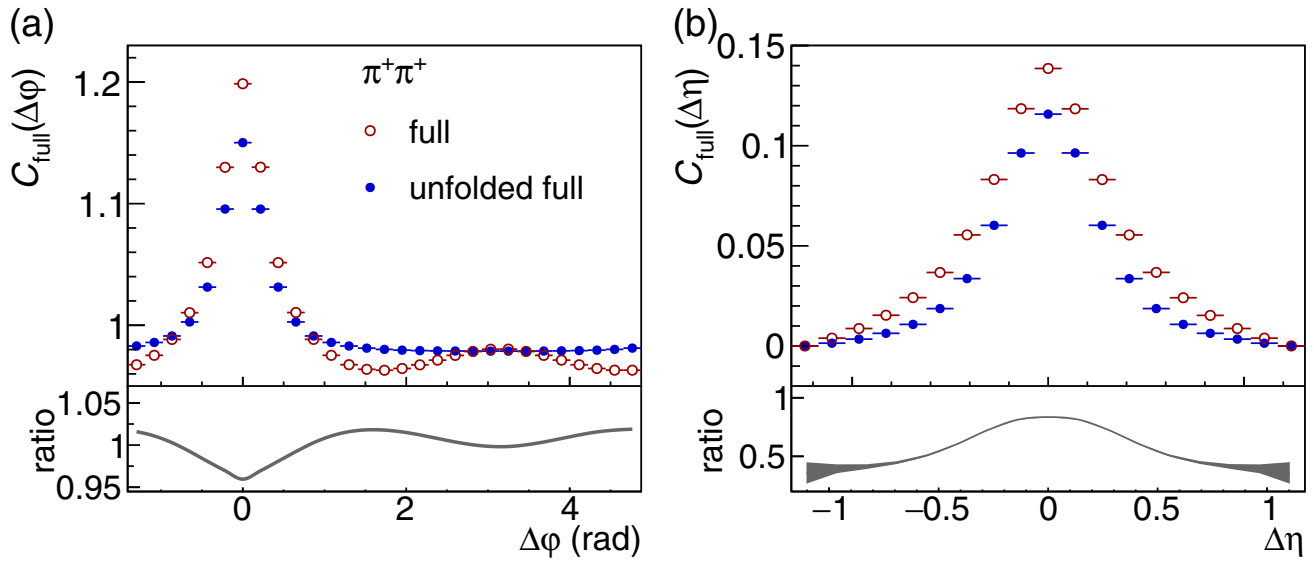

FIG. 7. Upper panels: projections of $C_{\mathrm{full}}(\Delta\eta, \Delta\varphi)$ (closed symbols) and $C^{\mathrm{unfolded}}_{\mathrm{full}}(\Delta\eta, \Delta\varphi)$ (open symbols) in $\Delta\varphi$ for $|\Delta\eta| < 1.3$ (left column) and in $\Delta\eta$ for $-\frac{\pi}{2} < \Delta\varphi < \frac{\pi}{2}$ (right column) for $\pi^+\pi^+$ pairs from PYTHIA simulated pp collisions at $\sqrt{s} = 13$ TeV. For better visualization of the differences, the $\Delta\eta$ projections are shifted by a constant value so that the baseline is at $C(\Delta\eta) = 0$. Lower panels: ratio of $C^{\mathrm{unfolded}}_{\mathrm{full}}$ to $C_{\mathrm{full}}$.

## ACKNOWLEDGMENTS

We would like to thank Prof. A. Kisiel for his invaluable suggestions. This work was supported by the Polish National Science Centre under Decisions No. UMO-2016/22/M/ST2/00176, No. UMO-2017/27/B/ST2/01947, and No. UMO-2021/41/B/ST2/03732, as well as by the IDUB-POB-FWEiTE-1 Project granted by the Warsaw University of Technology under the Program Excellence Initiative: Research University (ID-UB).

## APPENDIX: LEDNICKÝ AND LYUBOSHITZ FORMALISM

The pair wave function in Eq. (2) $\Psi(\mathbf{k}^*, \mathbf{r}^*)$ depends on the interaction between the two particles in the pair. In the most general case, the strong and Coulomb forces have to be taken into account. In such a scenario, the interaction of two (non-identical) particles is given by the Bethe-Salpeter amplitude, which is a solution of the quantum scattering problem taken with the inverse time direction (which is denoted by the "−" sign in front of $k^*$),

$$\Psi^{(+)}_{-\mathbf{k}^*}(\mathbf{r}^*) = e^{i\delta_c}\sqrt{A_C(\eta)}\left[e^{-i\mathbf{k}^*\cdot\mathbf{r}^*}F(-i\eta, 1, i\zeta^+) + f_C(\mathbf{k}^*)\frac{\tilde{G}(\rho, \eta)}{r^*}\right], \quad \text{(A1)}$$

where $\delta_c = \arg\Gamma(1 + i\eta)$ is the Coulomb $s$-wave phase shift, $A_c(\eta) = 2\pi\eta(e^{2\pi\eta} - 1)^{-1}$ is the Gamow factor (Coulomb penetration factor), $\zeta^\pm = k^*r^*(1 \pm \cos\theta^*)$, $\eta = 1/(k^*a_C)$, $F$ is the confluent hypergeometric function, and $\tilde{G}$ is the combination of the regular and singular $s$-wave Coulomb functions. Symbol $\theta^*$ denotes the angle between the pair relative momentum and relative position in the pair rest frame, whereas $a_C$ is the Bohr radius of the pair. The component $f_C(k^*)$ is the strong interaction scattering amplitude, modified by the Coulomb component,

$$f_C^{-1}(k^*) = \frac{1}{f_0} + \frac{1}{2}d_0k^{*2} - \frac{2}{a_C}h(k^*a_C) - ik^*a_C, \quad \text{(A2)}$$

where $a_C$ is the Bohr radius, $\zeta = k^*r^*(1 + \cos\theta^*)$, $\theta^*$ is the angle between $\vec{k}^*$ and $\vec{r}^*$, $\eta = 1/(k^*a_C)$, $F$ is the confluent hypergeometric function, $G$ combines the regular and singular Coulomb functions, and $h(\eta) = \eta^2\sum_{n=1}^{\infty}[n(n^2 + \eta^2)]^{-1} - \gamma - \ln|\eta|$ ($\gamma = 0.5772$ is the Euler constant).

In addition, if identical particles are considered, it has to be properly symmetrized (antisymmetrized) for pairs of identical bosons (fermions). Moreover, the description becomes more complicated when coupled channels are present. For details see Refs. [17,35].